\documentclass[letterpaper, 10 pt, conference]{IEEEtran}  

\IEEEoverridecommandlockouts                              
\newcommand\Mark[1]{\textsuperscript{#1}}

\usepackage[switch]{lineno} 

\usepackage{lettrine}

\usepackage{graphicx} 
\usepackage{url}
\usepackage{cite}

\title{\LARGE \bf  Fast and Reproducible LOFAR Workflows with AGLOW\\[.75ex] 
  {\normalfont\large 
  A.P. Mechev\Mark{a,1},J.B.R Oonk\Mark{a,b,c,2}, T. Shimwell\Mark{b,3}, A. Plaat\Mark{d,4}, H.T. Intema\Mark{a,5}, H.J.A. R\"{o}ttgering\Mark{a,6}
  }\\[-1.5ex]
}

\author{

\IEEEauthorblockA{%
        \Mark{a}Leiden Observatory,\\
        Leiden University,\\
        Niels Bohrweg 2 NL-2333 CA\\
        Netherlands \\
       \Mark{1}apmechev at strw.leidenuniv.nl \\
       \Mark{2}oonk at strw.leidenuniv.nl \\
       \Mark{5}intema at strw.leidenuniv.nl \\
       \Mark{6}rottgering at strw.leidenuniv.nl \\
    }  \and

\IEEEauthorblockA{%
        \Mark{b}ASTRON,\\
        Oude Hoogeveensedijk 4,\\
       7991 PD Dwingeloo\\
        Netherlands\\
        \Mark{3}shimwell at astron.nl
    } \and
\IEEEauthorblockA{%
        \Mark{c}SURFsara\\
        P.O. Box 94613\\
        1090 GP Amsterdam\\
        Netherlands\\   
        }  \and
\IEEEauthorblockA{%
        \Mark{d}LIACS,\\
        Leiden University,\\
        Niels Bohrweg 1 NL-2333 CA\\
        Netherlands\\
        \Mark{4}a.plaat at liacs.leidenuniv.nl
    }
}

\usepackage[perpage]{footmisc}
\begin{document}

\maketitle
\thispagestyle{empty}
\pagestyle{empty}

\begin{abstract}

The LOFAR radio telescope creates Petabytes of data per year. This data is important for many scientific projects. The data needs to be efficiently processed within the timespan of these projects in order to maximize the scientific impact. We present a workflow orchestration system that integrates LOFAR processing with a distributed computing platform. The system is named Automated Grid-enabled LOFAR Workflows (AGLOW). AGLOW makes it fast and easy to develop, test and deploy complex LOFAR workflows, and to accelerate them on a distributed cluster architecture. AGLOW provides a significant reduction in time for setting up complex workflows: typically, from months to days. We lay out two case studies that process the data from the LOFAR Surveys Key Science Project. We have implemented these into the AGLOW environment. We also describe the capabilities of AGLOW, paving the way for use by other LOFAR science cases. In the future, AGLOW will automatically produce multiple science products from a single dataset, serving several of the LOFAR Key Science Projects.

Target:  IEEE International Conference On e-Science
Keywords: Workflow management software, Radio Astronomy, Distributed Computing, Big Data applications
\end{abstract}

\renewcommand{\thefootnote}{\fnsymbol{footnote}}
\section{Introduction}\label{sec:intro}

Data sets in radio astronomy have increased 1000-fold over the past decade\cite{sabater_datasize}. It is no longer feasible to move, store and process these data sizes at university clusters, nor to process these data manually. LOFAR, the Low-Frequency Array\cite{LOFAR} is a modern and powerful radio telescope that creates more than 5 Petabytes of data per year. At present, the majority of LOFAR time is allocated to several Key Science Projects (KSPs)\cite{lotss}. These projects need to process hundreds or thousands of observations. Typical observations produce approximately 14 TB of archived data. Obtaining high fidelity images from this data requires complex processing steps. To manage and automate the data processing, workflow management software is needed. This software needs to accelerate LOFAR processing on a High Throughput Computing (HTC) cluster while ensuring it is easy to prototype, test, and integrate future algorithms and pipelines. 
	
To automate LOFAR data processing, we have worked with the LOFAR Surveys KSP (SKSP). Together, we designed a software suite that integrates LOFAR software\cite{cookbook} with the Dutch grid infrastructure\cite{dutchinfra}. This software, based on Apache Airflow\footnote{\url{https://airflow.apache.org/}}, makes it easy to add future science cases, extend and modify pipelines, include data quality checks, and rapidly prototype complex pipelines.  For the SKSP use cases, AGLOW achieves a significant reduction in development time: from months to days, allowing researchers to concentrate on data analysis rather than management of processing.  Additionally, and perhaps more importantly, the software versions and repositories used are defined within the workflow. This makes reproducibility an integral part of the AGLOW software.  Finally, the software is built to  leverage an HTC cluster by seamlessly submitting the processing jobs through the cluster's job submission system\cite{glite}. The work presented here builds on our previous work parallelizing single LOFAR jobs\cite{mechev} on a distributed environment. The majority of processing was done at SURFsara at the Amsterdam Science Park\cite{SurfSara}, which is one of the sites used by the LOFAR Long Term Archive (LTA)\footnote{\url{https://lta.lofar.eu}}. Ongoing efforts include scheduling and processing data at clusters in Pozna\`{n} in Poland and J\"{u}lich in Germany. 

\textbf{Contributions:}
The main features of the AGLOW software are the following:
\begin{itemize}
\item Integration of the Grid middleware with Apache Airflow, allowing us to dynamically define, create, submit and monitor jobs on the Dutch national e-infrastructure.
\item Integration of the LOFAR (LTA) utilities in  Airflow, facilitating pipeline developers to automate staging (moving from tape to disk) and retrieval of LOFAR data.
\item Integration of the SURFsara storage with Airflow, making LOFAR pipelines aware of the storage layer available at the Dutch national e-infrastructure.  
\item Ease of creating simple software blocks, with which users can integrate and test their pipelines. 
\item Storing all software versions and script repositories as part of the workflow to make LOFAR processing reproducible and portable. 
\end{itemize}

\textbf{Outline:}
The organization of this manuscript is as follows: We provide background on data processing in radio astronomy and why LOFAR science requires complex workflows and cover workflow management algorithms and capabilities (section \ref{sec:background}). We discuss related work in workflow management (section \ref{sec:related}). In section \ref{sec:AGLOW}, we introduce our software and two use cases. Both of our use cases require acceleration at an HTC cluster and automation by a workflow orchestration software. We follow these examples with details on the integration between LOFAR software, LOFAR data and the resources at SURFsara in Amsterdam in section \ref{sec:extending}. Finally, we discuss our results (sect. \ref{sec:results}) and look ahead to the demands of future LOFAR projects and upcoming telescopes in section \ref{sec:conclusions}. 

\section{Background}\label{sec:background}

This work lies at the intersection of Radio Astronomy and Computer Science. The goal of the study is to leverage the flexibility of an industry standard workflow management software and use CERN's Worldwide Computing Grid\footnote{\url{http://wlcg.web.cern.ch/}} at SURFsara \cite{grid} to accelerate reproducible processing of LOFAR data.

A single LOFAR surveys observation is recorded in distinct frequency chunks (henceforth called `subbands'), each of which is uploaded to the LTA as a separate file. Some of the processing steps require the entire frequency information, while others can run independently and operate on a single subband. The latter steps can be easily accelerated on an HTC cluster by taking advantage of the data level parallelism. 

Multiple scientific projects may desire to run different processing steps on a single LOFAR observation. To minimize time spent on retrieving data from the LTA and eliminate re-processing of data, pipelines for multiple science cases need to be integrated together. This integration should be done by a software that encodes the dependencies between different steps and automatically executes processing steps once their dependencies have been met. Software packages that solve these challenges are called `workflow management software' (see, e.g., \cite{workflow1,workflow2,workflow3}.)

\section{Related Work}\label{sec:related}

A workflow is described by a set of tasks. The dependencies between these tasks are encoded in a Directed Acyclic Graph (DAG) \cite{dag}. This data structure imposes a strict dependency hierarchy between the tasks \cite{dagalgos}. This means that there exists a well-defined execution order and a well-defined list of dependencies for each task. The execution order is typically determined by algorithms such as Kahn's algorithm \cite{Kahn} or a depth-first search \cite{dfs}.

Workflow management software is used in various fields from research to industry. In biology, gene sequencing and analysis pipelines require automation of multiple processing steps. In gene sequencing, Toil\footnote{\url{https://toil.readthedocs.io}} has been successfully used to automate RNA sequence analysis\cite{toil}. Additionally, many  software teams in biotech develop their own in-house workflow management software \cite{nextflow}.

Currently, we can parallelize a single processing step of the pipeline using the Grid LOFAR Tools (GRID\_LRT\footnote{\url{https://github.com/apmechev/GRID_LRT}}) \cite{mechev}. The LOFAR Surveys science cases incorporate multiple steps with inter-linked dependencies. Resolving these dependencies can be done efficiently by a comprehensive workflow orchestration software. The purpose of such software is to resolve dependencies between the multiple tasks in a workflow, execute these tasks, and track the status, logs, output, and runtime of each task. 

In astronomy, workflow systems have been developed that are telescope specific, such as ESOReflex\cite{reflex} by the European Southern Observatory. Other projects, such as astrogrid\footnote{http://www.astrogrid.org} and 'Workflow 4Ever'\footnote{http://wf4ever.github.io/ro/}, have either been completed or are no longer supported. The astrogrid project, for example, was a collaboration to create standards, infrastructure, and software for distributed astronomical processing. Its operation phase spanned 2008-2010. Workflow4Ever, likewise, has been out of support since 2013. To ensure continuing support for the LOFAR workflows, we have decided to use a leading enterprise workflow management software, Airflow.

Airflow is an open source Python software package developed by Airbnb\footnote{\url{https://www.airbnb.com/}} to manage complex workflows. It encodes workflows in Python and makes it easy to re-use, re-arrange, schedule and execute blocks in a user-defined workflow. Airflow is capable of scheduling and executing workflows by resolving the dependencies between tasks and scheduling these tasks for execution. The software uses a metadata database\footnote{In our case implemented by Postgresql} to retain metadata such as task state, execution date, and output. While Airflow allows building workflows easily from Python and bash functions, it can easily be extended to support custom processing scenarios. Additionally, Airflow conforms to the Common Workflow Language (CWL) \cite{cwl} standard using the \textit{cwl-airflow} package \cite{cwlairflow}, meaning it can execute CWL workflows as well. Finally, Airflow is part of the Apache incubator and upon certification will receive continual support by the Apache software foundation\footnote{https://www.apache.org/}.

\section{AGLOW}\label{sec:AGLOW}

Complex astronomical pipelines are time consuming to develop and operate. Furthermore, they may evolve rapidly to incorporate new processing techniques or requirements. Migrating these pipelines to a distributed, high throughput environment is often justified, or even required, in order to meet the timelines set by scientific projects. The time saved by running on a cluster must be balanced by the flexibility and development time required to implement or update the scientific pipelines. To address these concerns, we have developed a software package, Automated Grid-enabled LOFAR Workflows (AGLOW)\footnote{\url{https://github.com/apmechev/AGLOW}}. AGLOW is based on Airflow and the LOFAR software and addresses issues with automation and acceleration of LOFAR processing. 

With AGLOW, we can translate LOFAR pipelines into DAGs. We provide the tools that enable users to easily implement their LOFAR science pipelines and execute them on a distributed architecture. Using these tools, the data processing required by various LOFAR science cases is automated and accelerated.

\subsection{AGLOW: Case Study}

For our case-study, we have chosen two ways to process LOFAR Surveys data: coverage and depth. The Surveys Key Science Project (SKSP)\cite{lotss} is an ambitious project to map the northern sky at low frequencies using the Dutch LOFAR stations. These maps will help understand the formation and evolution of massive black holes, galaxies, clusters of galaxies and large-scale structure of the Universe.  

The LOFAR surveys observations consist of several tiers with the widest Tier (Tier 1) covering the whole sky visible from the Northern Hemisphere with 3168 observations of 8 hours each \cite{lotss}. The other tiers (Tier 2, Tier 3) consist of much longer observations of smaller sections of the sky and can collect hundreds of hours of data for a single  direction. The deepest single field being analyzed, in collaboration with the EoR group, is the North Celestial Pole (NCP) field which has $\sim$1700 hrs of observations to date.  Processing this data will create an image with an unprecedented resolution and sensitivity. Here we have implemented processing pipelines for both the Tier 1 data and Tier 2 data into AGLOW. 

The scientific importance of these two examples, as well as the large processing requirements, make them ideal candidates for acceleration and automation with AGLOW.

\subsubsection{Surveys Project: All Sky Survey}\label{sec:coverage}

The main driver for the development of AGLOW and its constituent packages has been the LOFAR SKSP Project. A typical 8-hour observation produces 14 TB of data. This data is eventually reduced to several hundred gigabytes. Data needs to be processed by two pipelines: first by the Direction Independent (DI) pipeline, and then by the Direction Dependent (DD) pipeline. 

We have split the DI pipeline into four stages, and the DD pipeline into two subsequent stages. Splitting up the pipelines in stages allows speedup through parallelization for the stages that can benefit from data-level parallelism. Additionally, this setup allows fault tolerance and easy re-processing. Current SKSP processing is easily started by launching a new DAG-run in Airflow. Importantly, with AGLOW, adding new functionality to the pipeline is easy and can be done at any time without disrupting current processing. 

\subsubsection{Deeper Surveys Fields}

To create deep images of a single field, minor modifications were made to the processing pipeline described in the previous section. Scripts were included to re-align the data in the frequency axis, and the DD processing steps include an extra final combination step that stacks multiple observations. Being able to rapidly test alternative processing strategies is crucial to creating a deep image of the NCP field. With the success of this project, future deep LOFAR observations will be processed with these pipelines.

\subsection{AGLOW: Implementation}\label{sec:implementation}

\begin{figure}[thpb]
 \centering
  \framebox{\parbox{3.3in}{      
  \includegraphics[width=0.45\textwidth]{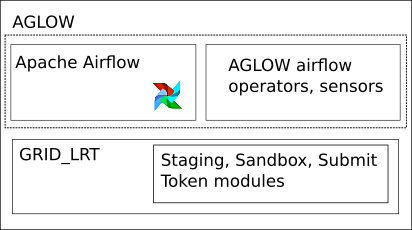}
 }}
 \caption{Design of the AGLOW software, incorporating Airflow, the GRID\_LRT package\cite{mechev} and custom operators designed to integrate LOFAR software, Grid middleware and dCache storage. GRID\_LRT is a software package developed to parallelize single LOFAR jobs at SURFsara. It contains several modules to help set up, and launch jobs on an HTC cluster at SURFsara. Airflow is a stand-alone package by Airbnb, which is extended with several classes that couple Airflow with the Grid infrastructure. These classes are collectively named the AGLOW operators/sensors. }
 \label{AGLOW_structure}
\end{figure}

AGLOW combines the LOFAR software, the Grid LOFAR Tools (GRID\_LRT), and Airflow to allow automation and makes large-scale LOFAR processing easily reproducible. The components of the AGLOW software are shown in Fig. \ref{AGLOW_structure}.  In this section, we will discuss these components and their functions. 

\subsubsection{GRID LOFAR Tools and LOFAR software}
We have previously developed tools to create LOFAR jobs and launch them on a distributed infrastructure\cite{mechev}. These tools have matured to a point where it is easy to both plug and play existing scripts and extend the framework to add more complex pipelines. These steps make it possible for a user to batch execute bash or Python scripts on their LOFAR data in parallel. After the scripts are executed, the results are uploaded to shared dCache storage\cite{dcache} at SURFsara\cite{SurfSara}. 

More complex steps use additional Github repositories, such as the prefactor\footnote{https://github.com/lofar-astron/prefactor} for direction independent calibration or DDF-pipeline\footnotemark{}  for direction-dependent calibration and imaging. The sequence of steps is encoded in parameter-set files (parsets), which can be modified and dropped into AGLOW depending on the processing requirements.

With AGLOW, we can easily include the DDF-pipeline and prefactor repositories, as well as any other scripts. Since these scripts are tracked by git \cite{torvalds2010git}, a full commit and branch history of the scripts is available. We use this history to make processing reproducible, by using the same git-commit for all LOFAR datasets. 

In addition to these script repositories, we have integrated the most common software packages used to process LOFAR data with AGLOW. These are the Default Pre-Processing Pipeline (DPPP)\cite{cookbook}, the LOFAR Solutions Tool (LoSoTo), WSclean \cite{wsclean}, AOflagger\cite{aoflag}, CASA\cite{casa}, pyBDSF\cite{bdsf}, DDFacet\cite{tasseDDFacet} and KillMS\cite{smirnov_tasse_KillMS,tassekalman}.

\footnotetext{https://github.com/mhardcastle/ddf-pipeline. DDF-pipeline is a leading example of a Direction Dependent calibration pipeline used for LOFAR data. It uses DDFacet \cite{tasseDDFacet}, KillMS \cite{smirnov_tasse_KillMS} and to create high quality images. }

\subsubsection{Extending Airflow}\label{sec:extending}

Two types of modifications were made to Airflow to allow processing on a Grid environment. First, functions were added to check the number of files located in intermediate grid storage. We use this to decide whether to stage files, or whether enough files have been successfully processed by a previous task. 

\begin{figure}[thpb]
 \centering
  \framebox{\parbox{3.3in}{      
  \includegraphics[width=0.45\textwidth]{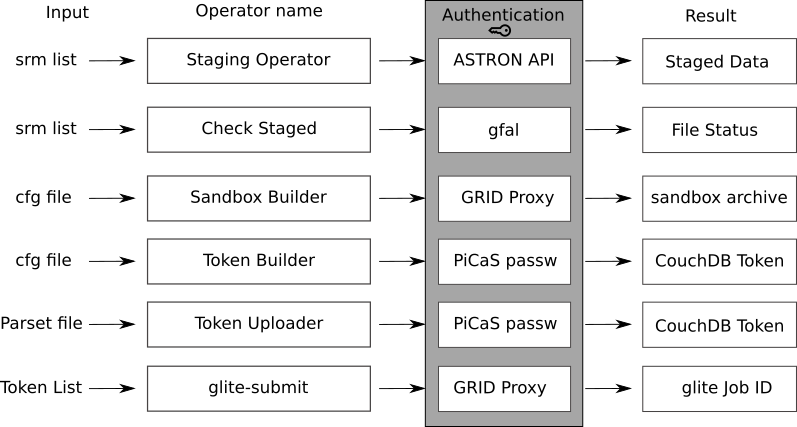}
 }}
 \caption{Airflow Operators for Staging LOFAR data, creating job descriptions and submitting jobs to the Dutch grid. On the left is the input given to each operator. `SRM lists' are lists of links to data at the LOFAR LTA or located on the SURFsara dCache storage. Parsets are files specific to `prefactor' and `DDF-pipeline' and define the processing for each pipeline step. Finally, the `Sandbox' and `Token' operators read their parameters from a configuration file. The use of a scripts sandbox and job description tokens is detailed in our previous work\cite{mechev}.}
 \label{AGLOW_Operators}
\end{figure}

Second, more complex tasks were implemented as Airflow operators or sensors (Figure \ref{AGLOW_Operators}). These tasks include creating job description files, setting up job scripts, launching jobs using the gLite workload management system and monitoring the status of these jobs. Future additions will include operators that evaluate the current cluster workload and make decisions on location to launch the data processing. With the AGLOW package, such tasks are easy to implement without modifying or interrupting processing. This leads to an easily reproducible, intelligent scientific processing that is also efficiently executed and requires minimal interaction. The operators and sensors added to Airflow are shown in Fig. \ref{AGLOW_Operators}.

Using AGLOW to accelerate the execution of a pipeline requires deciding how to split the processing to benefit from parallelization. Once the steps to be parallelized are selected, users can add git repositories of scripts to the configuration file. Next, each step is added to a Python script called the DAG file. This file is placed in the Airflow's dags folder, which adds it to AGLOW. To migrate LOFAR workflows to a new server, the DAG and configuration files need to be transferred to the new AGLOW instance.

\subsection{AGLOW: Jobs}

Once LOFAR observations are downloaded from the LTA, they are typically processed with several packages before producing a science ready dataset. We have integrated these packages with Airflow to make it easy to create complex LOFAR workflows. 

Each of the processing steps above requires extra set-up to process on the Dutch Grid infrastructure. The job scripts setup, job description, and job submission are done by the GRID\_LRT package\cite{mechev}. With AGLOW, we automate this setup, enabling users to focus on developing more comprehensive data processing pipelines. Below we outline several possible steps a user can use in their pipeline. 

\subsubsection{DPPP Parset}
The DPPP software is used extensively in LOFAR data processing. It has many capabilities such as flagging bad data, averaging data in time and frequency, and calibrating the data with a sky-model. 

The input parameters of this software are stored in a text file called a parset. The input data and the DPPP parset are sufficient to define a DPPP execution step. As noted in section \ref{sec:background}, LOFAR data is split in frequency into subbands. Much of the DPPP processing, such as averaging and flagging, can be done independently for each subband, thus they can be processed on independent machines. This parallelization makes these steps a perfect candidate for an HTC cluster. For a dataset that is split into 244 subbands, 244 jobs are launched concurrently. 

\begin{figure}[thpb]
 \centering
 \framebox{\parbox{2.5in}{      
  \includegraphics[width=0.35\textwidth]{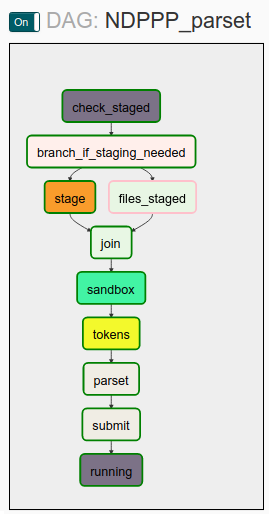}
 }}
  \caption{Render of the DPPP parset DAG in the Airflow User Interface. This view shows the setup and submission steps. Even this simple DAG can include branching options such as the \texttt{branch\_if\_staging\_needed} task which checks if the data is not staged and stages it. All of the operators in this figure are part of the AGLOW software. Their inputs and outputs are shown in Fig. \ref{AGLOW_Operators}. Using configuration files, the NDPPP DAG can be used by different users for different science cases making it portable and maintainable. These features make reproducible science with LOFAR data easy. 
 }
 \label{fig:NDPPP_DAG}
\end{figure}

In Airflow, the DPPP parset task is encoded in a DAG (Fig. \ref{fig:NDPPP_DAG}). The DPPP DAG is a linear workflow that consists of the 'sandbox' setup, creation of the job-description documents, uploading of the DPPP parset and job launching and monitoring.

\subsubsection{WSclean Job}

The WSClean\cite{wsclean} package is used to create an image from a LOFAR dataset. This software has a very wide range of parameters options, however, it cannot take a parset file as an input. Instead, the parameters are specified in the command line. In the AGLOW implementation, we parse all the command-line parameters from a text file, referred to as the 'wsclean parset'. This file is added to the jobs in the same way as the DPPP parset, i.e. using the \textit{Token Uploader} Operator. The DAG for the wsclean software uses the same blocks as the DPPP DAG, with different configuration and parset files. The reuse of Airflow operators makes maintainability of all tasks easier. 

\subsection{Shell/Python Script}

Users that require the run of multiple software packages on a single dataset can craft a custom shell or Python script that executes these steps using the LOFAR tools during a single distributed job. This option increases flexibility and minimizes the overhead associated with scheduling and running multiple jobs in sequence. At the workflow orchestration level, we use the same Airflow operators as the above tasks. The script is uploaded to the job description database using the \textit{Token Uploader} Operator. It is executed once the jobs are launched. 

Currently only the LOFAR Spectroscopy project uses custom shell scripts to process LOFAR data. A recent study of carbon recombination lines used a custom bash script to calibrate and image LOFAR data on the SURFsara \texttt{GINA}\footnote{The \texttt{GINA} cluster is an HTC cluster located at SURFsara integrated with the Dutch Grid initiative. It supports massively parallel processing which is required to efficiently process LOFAR data with \textit{prefactor}. } cluster \cite{pedro}. 

\subsection{Prefactor parset}

The input to the prefactor pipeline software is a parset file which describes a linear workflow. The description of this workflow consists of a list of processing steps and their associated parameters. The `prefactor' package uses the LOFAR software to do the direction-independent calibration of the archived LOFAR datasets. Prefactor steps are executed by the generic pipeline framework\cite{cookbook}. While this framework can run a sequential pipeline, it is not capable of conditional branching nor parallelization on all cluster architectures. The original goal of the GRID\_LRT software was to tackle the parallelization challenge while AGLOW solves the additional challenge of pipeline management. 

We have already processed more than 50 datasets through the `prefactor' DAG using AGLOW. The full `prefactor' pipeline is shown in figure \ref{SKSP_workflow}. This DAG shows the four processing steps as well as additional Python operators that manage the staging and result verification. 

\subsection{DDF-pipeline}

The final AGLOW DAG is the implementation of the DDF-pipeline repository which is a pipeline that is extensively used by the LOFAR surveys KSP and is described in detail in \cite{lotss}. This pipeline operates on the products of the prefactor pipeline and consists of a series of calibration and imaging loops with the objective of creating a final science quality image. For each of these loops the majority of the processing time is spent in DDFacet\cite{tasseDDFacet} and KillMS\cite{smirnov_tasse_KillMS,tassekalman} steps that perform the direction-dependent imaging and calibration respectively. 

In total, DDF-pipeline takes $\sim$4\, days of processing to complete. As DDF-pipeline creates large intermediate files we have so far not divided the pipeline into too many steps to avoid filling the storage on the \texttt{GINA} cluster. However, we have split the pipeline into two steps and there is further potential for parallelization that will be implemented in the future. 

\subsection{Linking Multiple Jobs}

Pre-processing of LOFAR SKSP data can be done by a single DPPP task, with 244 jobs running in parallel. More complex LOFAR pipelines will include multiple processing tasks as well as tasks responsible for job setup. Therefore, it is important to facilitate running multi-step pipelines with AGLOW. 

Creating workflows by defining dependencies between tasks is a core Airflow capability. We use this functionality to link multiple steps of a LOFAR pipeline together. In the SKSP pipeline, we take advantage of the data level parallelism for the initial processing steps for the calibrator and target. The other two  steps are run as a single grid job. Switching the parallelization for each step is done by changing the number of datasets per node parameter in the configuration file for each step. 

\section{Results and Discussions}\label{sec:results}

The implementation of AGLOW makes it possible to efficiently process LOFAR data with minimal user interaction. The scheduling algorithm automatically launches pipelines, meaning that there is little time spent between runs. Additionally, controlling/fixing the version of the scripts is done by specifying the commit of each script repository. This makes data processing easily reproducible. Once the dependencies of multiple science pipelines have been encoded in a DAG, Airflow efficiently executes this DAG, running tasks in parallel where possible. 

The first LOFAR processing pipeline integrated with AGLOW was a single linear workflow, with only one submission to the compute cluster. This workflow is used to reduce the data size making data retrieval to research institutes less time consuming. We offer this workflow as a service to LOFAR users who do not have a high-bandwidth connection to the LOFAR Archive. 

A more complex pipeline was implemented: the LOFAR direction independent calibration pipeline (`prefactor'). The scientific importance and complexity of this pipeline make it a good case study for the capabilities of the AGLOW software. We show that AGLOW's design allows integration of more complex data processing workflows with the Dutch Grid resources. These workflows can be either used by PIs of LOFAR projects or offered as a processing service to the wider astronomical community.

An important feature of AGLOW is the loose coupling between pipeline logic, software versions, pipeline parameters, and datasets. The goal of this decoupling is to give users complete control over all the processing variables. With AGLOW, one can develop the pipeline logic independently of the LOFAR software versions and conversely update the LOFAR software and script repositories independently from the pipeline logic. Finally, the Airflow operators are themselves decoupled from the scientific pipelines. As these operators are reused, this decoupling makes them easy to maintain and extend.

In large part thanks to their flexibility, automation, and Grid integration, AGLOW and GRID\_LRT have become a standard part of the Direction Independent processing for the LOFAR SKSP project.

\section{Conclusions}\label{sec:conclusions}

In this work, we have detailed a comprehensive workflow management software for processing radio astronomy data on a distributed infrastructure. We leverage an industry standard workflow management software, Airflow. Using its capabilities, we make it possible to build, test, automate and deploy LOFAR pipelines on short timescales, generally from months to days. With the flexibility of Airflow's Python and Bash operators, users can design their own workflows, as well as co-ordinate more complex science cases. In this way, AGLOW facilitates reproducible processing of scientific data. In the future, AGLOW will support additional LOFAR science cases including Long Baselines and Spectroscopy. In this article, we have described our implementation of the data processing pipelines used by the LOFAR Surveys Key Science Project. 

Future work includes further de-coupling of the Grid-setup and pipeline logic. We will do this by creating `sub-dags' (details in \ref{sec:Subdags}) for each type of LOFAR jobs. Using these sub-dags will reduce the complexity of scientific workflows while also making the code even more reusable and thus easier to maintain and upgrade. Efforts to integrate processing at the other two LTA sites, (J\"{u}lich and Pozna\'{n}) have already started with `prefactor' runs being performed on J\"{u}lich using a modified version of the SKSP workflow. The software also currently works on the Eagle cluster at Pozna\'{n}. Combining the J\"{u}lich and SURFsara workflows will be done in the future so that AGLOW can track and start processing at multiple clusters. 

Finally, AGLOW can be used as a `LOFAR As A Service' model.  In this model, users only provide an observation ID and processing parameters and receive the final results upon job completion. This model will build upon previous success offering LOFAR processing to users without login to the \texttt{GINA} cluster \cite{oonk_prep}. This previous work was already useful for studying radio absorption in Cassiopeia A \cite{maria} and a `data-to-images' service will be valuable to the whole LOFAR community.

Our experience with automating LOFAR scientific workflows on a distributed architecture will be valuable when setting up data processing for future Radio Telescopes such as the Square Kilometer Array \cite{ska} .




 \section*{APPENDIX}
 
\begin{figure*}[!ht]
 \centering
 \includegraphics[width=0.52\textwidth]{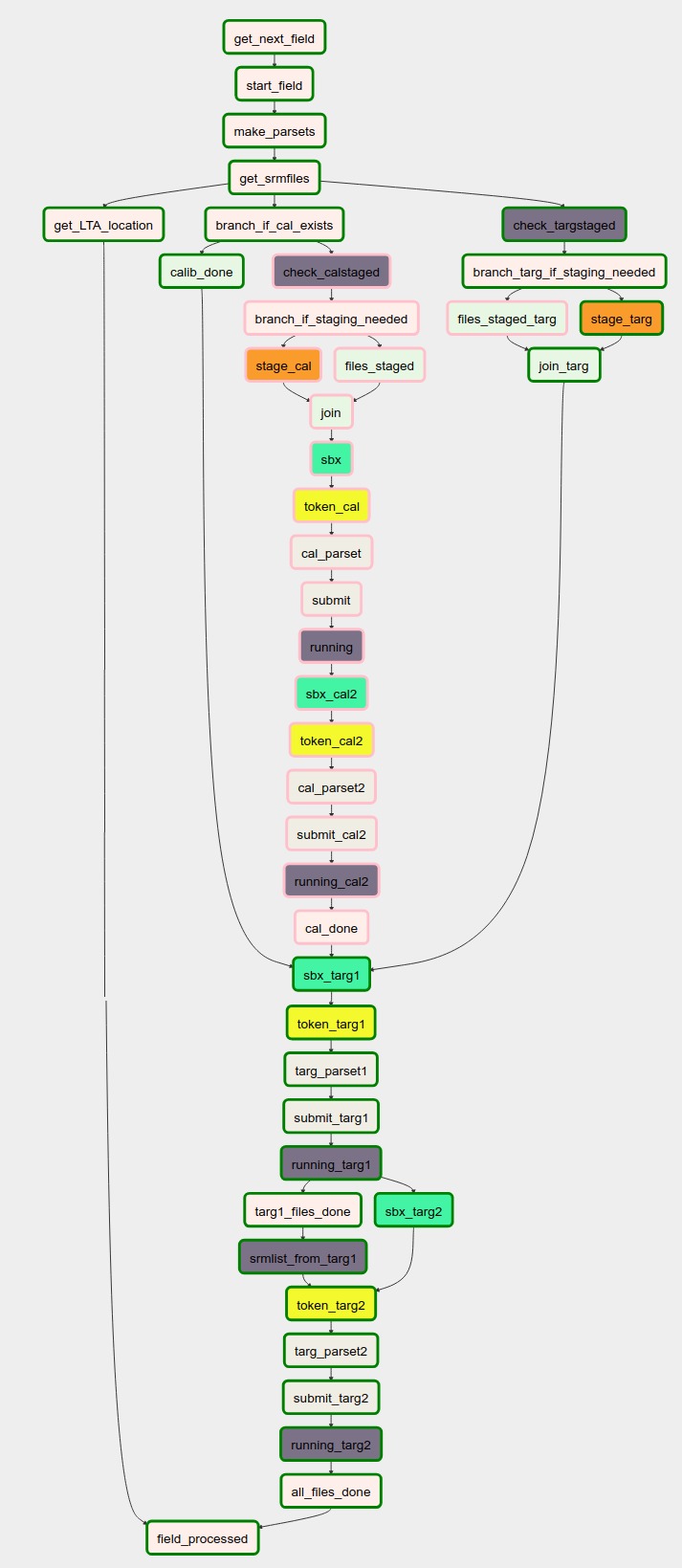}
 \caption{Workflow for the prefactor pipeline. Here we show the reuse of AGLOW operators for the four prefactor steps. In addition to the LOFAR processing, we also have conditional operators to skip processing of the calibrator if it has been previously processed. This is done by the `branch\_if\_cal\_exists' task. We also have a final step that checks if all the results have been uploaded, done by the `all\_files\_done' task. Likewise, quality checks can be added in this workflow wherever needed.}
 \label{SKSP_workflow}
\end{figure*}

The LOFAR SKSP workflow is shown in Figure \ref{SKSP_workflow}. This figure shows how reuse of the staging, setup operators, and glite-wms sensors makes maintainability easy and allows rapid prototyping of complex pipelines. 
 
 This workflow additionally takes advantage of Airflow's \textit{PythonOperator} to check if the LOFAR data is on disk at the archive and whether all final products were uploaded by each step. AGLOW also allows for staging the calibrator and target files concurrently. When the data is staged, Airflow continues with the processing of that data. 

\subsection{Sub-DAG}\label{sec:Subdags}
Airflow allows developers to include entire DAGs as a single task in their workflow. Airflow can trigger a DAG execution based on parameters provided by the parent DAG. This feature makes it possible to concatenate short, commonly used tasks into DAGs and call them in a parent workflow. Using sub-DAGS makes the code more maintainable and easy to use, while it makes workflows simpler. For LOFAR, Sub-DAGs are used to automate job submission, making the resulting scientific workflows simpler.


\section*{Acknowledgement}

APM would like to acknowledge the support from the NWO/DOME/IBM programme ``Big Bang Big Data: Innovating ICT as a Driver For Astronomy'', project \#628.002.001.

The processing and storage functionality that has made this project possible was enabled by SURF Cooperative through grant e-infra 160022 \& 160152. The LOFAR software and dedicated reduction packages
on \url{https://github.com/apmechev/GRID_LRT} were deployed on the e-infrastructure by the LOFAR e-infra group, consisting of J.B.R. Oonk (SURFsara \& Leiden Observatory), A.P. Mechev (Leiden Observatory).

This paper is based on data obtained with the International LOFAR Telescope (ILT) under project codes LC2\_038 and LC3\_008. LOFAR (van Haarlem et al. 2013) is the Low-Frequency Array designed and constructed by ASTRON. It has observing, data processing, and data storage facilities in several countries, which are owned by various parties (each with their own funding sources), and which are collectively operated by the ILT foundation under a joint scientific policy. The ILT resources have benefited from the following recent major funding sources: CNRS-INSU, Observatoire de Paris and Universit\'{e} d’Orl\'{e}ans, France; BMBF, MIWF-NRW, MPG, Germany; Science Foundation Ireland (SFI), Department of Business, Enterprise and Innovation (DBEI), Ireland; NWO, The Netherlands; The Science and Technology Facilities Council (STFC), UK

\bibliographystyle{unsrt}
\bibliography{bibliography}

\begin{thebibliography}{10}

\bibitem{sabater_datasize}
J~Sabater, S~S{\'a}nchez-Exp{\'o}sito, J~Garrido, JE~Ruiz, PN~Best, and
  L~Verdes-Montenegro.
\newblock Calibration of radio-astronomical data on the cloud. {LOFAR}, the
  pathway to {SKA}.
\newblock In {\em Highlights of Spanish Astrophysics VIII}, pages 840--843,
  2015.

\bibitem{LOFAR}
MP~Van~Haarlem, MW~Wise, AW~Gunst, George Heald, JP~McKean, JWT Hessels,
  AG~De~Bruyn, Ronald Nijboer, John Swinbank, Richard Fallows, et~al.
\newblock {L}{O}{F}{A}{R}: The low-frequency array.
\newblock {\em Astronomy \& astrophysics}, 556:A2, 2013.

\bibitem{lotss}
TW~Shimwell, HJA R{\"o}ttgering, Philip~N Best, WL~Williams, TJ~Dijkema,
  F~De~Gasperin, MJ~Hardcastle, GH~Heald, DN~Hoang, A~Horneffer, et~al.
\newblock The {L}{O}{F}{A}{R} {T}wo-metre {S}ky {S}urvey-{I}. {S}urvey
  description and preliminary data release.
\newblock {\em Astronomy \& Astrophysics}, 598:A104, 2017.

\bibitem{cookbook}
Dijkma~Tammo Jan.
\newblock Lofar imaging cookbook.
\newblock Available at
  \url{http://www.astron.nl/sites/astron.nl/files/cms/lofar_imaging_cookbook_v19.pdf
  }.

\bibitem{dutchinfra}
Jeff Templon and Jan Bot.
\newblock The dutch national e-infrastructure.
\newblock In {\em International Symposium on Grids and Clouds (ISGC)},
  volume~13, 2016.

\bibitem{glite}
Erwin Laure, A~Edlund, F~Pacini, P~Buncic, S~Beco, F~Prelz, A~Di~Meglio,
  O~Mulmo, M~Barroso, Peter~Z Kunszt, et~al.
\newblock Middleware for the next generation grid infrastructure.
\newblock Technical report, CERN, 2004.

\bibitem{mechev}
A.~{Mechev}, J.~B.~R. {Oonk}, A.~{Danezi}, T.~W. {Shimwell}, C.~{Schrijvers},
  H.~{Intema}, A.~{Plaat}, and H.~J.~A. {Rottgering}.
\newblock {An Automated Scalable Framework for Distributing Radio Astronomy
  Processing Across Clusters and Clouds}.
\newblock In {\em Proceedings of the International Symposium on Grids and
  Clouds (ISGC) 2017, held 5-10 March, 2017 at Academia Sinica, Taipei, Taiwan
  (ISGC2017). Online at
  \url{https://pos.sissa.it/cgi-bin/reader/conf.cgi?confid=293}, id.2}, page~2,
  March 2017.

\bibitem{SurfSara}
SURF.
\newblock Grid at {S}{U}{R}{F}sara.
\newblock \url{https://www.surf.nl/en/services-and-products/grid/index.html },
  2018.

\bibitem{grid}
Jamie Shiers.
\newblock The worldwide lhc computing grid (worldwide lcg).
\newblock {\em Computer physics communications}, 177(1-2):219--223, 2007.

\bibitem{workflow1}
Ilkay Altintas, Chad Berkley, Efrat Jaeger, Matthew Jones, Bertram Ludascher,
  and Steve Mock.
\newblock Kepler: an extensible system for design and execution of scientific
  workflows.
\newblock In {\em Scientific and Statistical Database Management, 2004.
  Proceedings. 16th International Conference on}, pages 423--424. IEEE, 2004.

\bibitem{workflow2}
David Churches, Gabor Gombas, Andrew Harrison, Jason Maassen, Craig Robinson,
  Matthew Shields, Ian Taylor, and Ian Wang.
\newblock Programming scientific and distributed workflow with triana services.
\newblock {\em Concurrency and Computation: Practice and Experience},
  18(10):1021--1037, 2006.

\bibitem{workflow3}
Ji~Liu, Esther Pacitti, Patrick Valduriez, and Marta Mattoso.
\newblock A survey of data-intensive scientific workflow management.
\newblock {\em Journal of Grid Computing}, 13(4):457--493, 2015.

\bibitem{dag}
Lin Wang.
\newblock Directed acyclic graph.
\newblock In {\em Encyclopedia of Systems Biology}, pages 574--574. Springer,
  2013.

\bibitem{dagalgos}
David~J Pearce and Paul~HJ Kelly.
\newblock A dynamic topological sort algorithm for directed acyclic graphs.
\newblock {\em Journal of Experimental Algorithmics (JEA)}, 11:1--7, 2007.

\bibitem{Kahn}
A.~B. Kahn.
\newblock Topological sorting of large networks.
\newblock {\em Commun. ACM}, 5(11):558--562, November 1962.

\bibitem{dfs}
Jianjun Zhou and Martin M{\"u}ller.
\newblock Depth-first discovery algorithm for incremental topological sorting
  of directed acyclic graphs.
\newblock {\em Information Processing Letters}, 88(4):195--200, 2003.

\bibitem{toil}
John Vivian, Arjun~Arkal Rao, Frank~Austin Nothaft, Christopher Ketchum, Joel
  Armstrong, Adam Novak, Jacob Pfeil, Jake Narkizian, Alden~D Deran, Audrey
  Musselman-Brown, et~al.
\newblock Toil enables reproducible, open source, big biomedical data analyses.
\newblock {\em Nature biotechnology}, 35(4):314, 2017.

\bibitem{nextflow}
Paolo Di~Tommaso, Maria Chatzou, Evan~W Floden, Pablo~Prieto Barja, Emilio
  Palumbo, and Cedric Notredame.
\newblock Nextflow enables reproducible computational workflows.
\newblock {\em Nature biotechnology}, 35(4):316--319, 2017.

\bibitem{reflex}
W.~{Freudling}, M.~{Romaniello}, D.~M. {Bramich}, P.~{Ballester}, V.~{Forchi},
  C.~E. {Garc{\'{\i}}a-Dabl{\'o}}, S.~{Moehler}, and M.~J. {Neeser}.
\newblock {Automated data reduction workflows for astronomy. The ESO Reflex
  environment}.
\newblock {\em Astronomy \& Astrophysics}, 559:A96, November 2013.

\bibitem{cwl}
Peter Amstutz, Michael~R Crusoe, Neboj{\v{s}}a Tijani{\'c}, Brad Chapman, John
  Chilton, Michael Heuer, Andrey Kartashov, Dan Leehr, Herv{\'e} M{\'e}nager,
  Maya Nedeljkovich, et~al.
\newblock Common workflow language, v1. 0.
\newblock 2016.

\bibitem{cwlairflow}
Michael Kotliar, Andrey Kartashov, and Artem Barski.
\newblock Cwl-airflow: a lightweight pipeline manager supporting common
  workflow language.
\newblock {\em bioRxiv}, page 249243, 2018.

\bibitem{dcache}
Patrick Fuhrmann and Volker G{\"u}lzow.
\newblock dcache, storage system for the future.
\newblock In {\em European Conference on Parallel Processing}, pages
  1106--1113. Springer, 2006.

\bibitem{torvalds2010git}
Linus Torvalds and Junio Hamano.
\newblock Git: Fast version control system.
\newblock {\em URL http://git-scm. com}, 2010.

\bibitem{wsclean}
A.~R. {Offringa}, B.~{McKinley}, N.~{Hurley-Walker}, F.~H. {Briggs}, R.~B.
  {Wayth}, D.~L. {Kaplan}, M.~E. {Bell}, L.~{Feng}, A.~R. {Neben}, J.~D.
  {Hughes}, J.~{Rhee}, T.~{Murphy}, N.~D.~R. {Bhat}, G.~{Bernardi}, J.~D.
  {Bowman}, R.~J. {Cappallo}, B.~E. {Corey}, A.~A. {Deshpande}, D.~{Emrich},
  A.~{Ewall-Wice}, B.~M. {Gaensler}, R.~{Goeke}, L.~J. {Greenhill}, B.~J.
  {Hazelton}, L.~{Hindson}, M.~{Johnston-Hollitt}, D.~C. {Jacobs}, J.~C.
  {Kasper}, E.~{Kratzenberg}, E.~{Lenc}, C.~J. {Lonsdale}, M.~J. {Lynch}, S.~R.
  {McWhirter}, D.~A. {Mitchell}, M.~F. {Morales}, E.~{Morgan},
  N.~{Kudryavtseva}, D.~{Oberoi}, S.~M. {Ord}, B.~{Pindor}, P.~{Procopio},
  T.~{Prabu}, J.~{Riding}, D.~A. {Roshi}, N.~U. {Shankar}, K.~S. {Srivani},
  R.~{Subrahmanyan}, S.~J. {Tingay}, M.~{Waterson}, R.~L. {Webster}, A.~R.
  {Whitney}, A.~{Williams}, and C.~L. {Williams}.
\newblock {WSCLEAN: an implementation of a fast, generic wide-field imager for
  radio astronomy}.
\newblock {\em Monthly Notices of the Royal Astronomical Society},
  444:606--619, October 2014.

\bibitem{aoflag}
A.~R. {Offringa}, J.~J. {van de Gronde}, and J.~B.~T.~M. {Roerdink}.
\newblock {A morphological algorithm for improving radio-frequency interference
  detection}.
\newblock {\em Astronomy \& astrophysics}, 539:A95, March 2012.

\bibitem{casa}
JP~McMullin, B~Waters, D~Schiebel, W~Young, and K~Golap.
\newblock Casa architecture and applications.
\newblock In {\em Astronomical data analysis software and systems XVI}, volume
  376, page 127, 2007.

\bibitem{bdsf}
Niruj Mohan and David Rafferty.
\newblock Pybdsf: Python blob detection and source finder.
\newblock {\em Astrophysics Source Code Library}, 2015.

\bibitem{tasseDDFacet}
C~Tasse, B~Hugo, M~Mirmont, O~Smirnov, M~Atemkeng, L~Bester, E~Bonnassieux,
  MJ~Hardcastle, R~Lakhoo, J~Girard, et~al.
\newblock Facetting for direction-dependent spectral deconvolution.
\newblock {\em arXiv preprint arXiv:1712.02078}, 2017.

\bibitem{smirnov_tasse_KillMS}
OM~Smirnov and Cyril Tasse.
\newblock Radio interferometric gain calibration as a complex optimization
  problem.
\newblock {\em Monthly Notices of the Royal Astronomical Society},
  449(3):2668--2684, 2015.

\bibitem{tassekalman}
Cyril Tasse.
\newblock Nonlinear kalman filters for calibration in radio interferometry.
\newblock {\em Astronomy \& Astrophysics}, 566:A127, 2014.

\bibitem{pedro}
P.~{Salas}, J.~B.~R. {Oonk}, R.~J. {van Weeren}, F.~{Salgado}, L.~K.
  {Morabito}, M.~C. {Toribio}, K.~{Emig}, H.~J.~A. {R{\"o}ttgering}, and
  A.~G.~G.~M. {Tielens}.
\newblock {{LOFAR} observations of decameter carbon radio recombination lines
  towards Cassiopeia A}.
\newblock {\em Monthly Notices of the Royal Astronomical Society},
  467:2274--2287, May 2017.

\bibitem{oonk_prep}
J.B.R. Oonk, A.P. Mechev, A~Danezi, C~Schrijvers, and T.W. Shimwell.
\newblock Radio astronomy on a distributed shared computing platform: The
  {LOFAR} case.

\bibitem{maria}
M~Arias, J~Vink, F~De~Gasperin, P~Salas, JBR Oonk, RJ~Van~Weeren,
  AS~Van~Amesfoort, J~Anderson, R~Beck, ME~Bell, et~al.
\newblock Low-frequency radio absorption in cassiopeia a.
\newblock {\em Astronomy \& Astrophysics}, 612:A110, 2018.

\bibitem{ska}
Peter~E Dewdney, Peter~J Hall, Richard~T Schilizzi, and T~Joseph~LW Lazio.
\newblock The square kilometre array.
\newblock {\em Proceedings of the IEEE}, 97(8):1482--1496, 2009.

\end{thebibliography}

\end{document}